\documentclass[floats,floatfix,showpacs,amssymb,prd,twocolumn,superscriptaddress,nofootinbib]{revtex4-1}

\usepackage{amssymb,amsmath,verbatim,mathtools,needspace,enumitem,etoolbox,graphicx,physics,microtype,afterpage,bm}

\usepackage[dvipsnames, usenames]{xcolor}

\definecolor{linkcolor}{rgb}{0.0,0.3,0.5}
\usepackage[unicode, colorlinks=true, linkcolor=linkcolor, citecolor=linkcolor, filecolor=linkcolor,urlcolor=linkcolor, pdfusetitle]{hyperref}
\usepackage[all]{hypcap}
\usepackage[T1]{fontenc}
\usepackage[utf8]{inputenc}
\usepackage{tabularx}
\usepackage{float}
\allowdisplaybreaks
\graphicspath{{./figure/}}

\newcommand{\llp}{\left [}
\newcommand{\rrp}{\right ]}

\newcommand{\lp}{\left (}
\newcommand{\rp}{\right )}

\newcommand{\co}{\text{\tiny cutoff}}
\def\ii{{\text{\tiny i}}}

\def\PBH{\text{\tiny  PBH}}

\newcommand{\be}{\begin{equation}\begin{aligned}}
\newcommand{\ee}{\end{aligned}\end{equation}}

\newcommand{\bbe}{\begin{align}}
\newcommand{\eee}{\end{align}}

\newcommand{\bea}{\begin{eqnarray}}
\newcommand{\eea}{\end{eqnarray}}

\def\beq{\begin{equation}}
\def\eeq{\end{equation}}

\def\d{{\rm d}}

\def\beqa{\begin{eqnarray}}

	\def\eeqa{\end{eqnarray}}
	
\def\lsim{\mathrel{\rlap{\lower4pt\hbox{\hskip0.5pt$\sim$}}
		\raise1pt\hbox{$<$}}}     %
\def\gsim{\mathrel{\rlap{\lower4pt\hbox{\hskip0.5pt$\sim$}}
		\raise1pt\hbox{$>$}}}     %

\def\d{{\rm d}}

\def\d{{\rm d}}

\def\PBH{\text{\tiny \rm PBH}}

\def\eeqa{\end{eqnarray}}

\def\bq{\begin{quote}}
\def\eq{\end{quote}}

 at 10truept

\def\eeqa{\end{eqnarray}}
\def\lsim{\mathrel{\rlap{\lower4pt\hbox{\hskip0.5pt$\sim$}}
  \raise1pt\hbox{$<$}}}     %
\def\gsim{\mathrel{\rlap{\lower4pt\hbox{\hskip0.5pt$\sim$}}
  \raise1pt\hbox{$>$}}}     %

\def\apj{\rm{ApJ}}
\def\apjl{\rm{ApJ}}

\def\mnras{\rm{MNRAS}}
\def\prd{\rm{PRD}}

\def\pasa{\rm{PASA}}
\def\pasp{\rm{PASP}}

\newcommand{\jhu}{\affiliation{Department of Physics and Astronomy, Johns Hopkins University, 3400 N. Charles
Street, Baltimore, MD 21218, USA}}

\definecolor{rb4}{HTML}{27408B}

\begin{document}

\title{
Constraining the primordial black hole scenario with Bayesian inference\\ and machine learning: the GWTC-2 gravitational wave catalog}

\author{Kaze W. K. Wong} 
\email{kazewong@jhu.edu}
\jhu

\author{Gabriele Franciolini}
\affiliation{D\'epartement de Physique Th\'eorique and Centre for Astroparticle Physics (CAP), Universit\'e de Gen\`eve, 24 quai E. Ansermet, CH-1211 Geneva, Switzerland}

\author{Valerio De Luca}
\affiliation{D\'epartement de Physique Th\'eorique and Centre for Astroparticle Physics (CAP), Universit\'e de Gen\`eve, 24 quai E. Ansermet, CH-1211 Geneva, Switzerland}

\author{Vishal Baibhav} 
\jhu

\author{Emanuele Berti} 
\jhu

\author{Paolo Pani}
\affiliation{Dipartimento di Fisica, Sapienza Università 
di Roma, Piazzale Aldo Moro 5, 00185, Roma, Italy}
\affiliation{INFN, Sezione di Roma, Piazzale Aldo Moro 2, 00185, Roma, Italy}

\author{Antonio~Riotto}
\affiliation{D\'epartement de Physique Th\'eorique and Centre for Astroparticle Physics (CAP), Universit\'e de Gen\`eve, 24 quai E. Ansermet, CH-1211 Geneva, Switzerland}
\affiliation{INFN, Sezione di Roma, Piazzale Aldo Moro 2, 00185, Roma, Italy}

\date{\today}

\begin{abstract}
\noindent
Primordial black holes~(PBHs) might be formed in the early Universe and could comprise at least a fraction of the dark matter. 
Using the recently released GWTC-2 dataset from the third observing run of the LIGO-Virgo Collaboration, we investigate whether current observations are compatible with the hypothesis that all black hole mergers detected so far are of primordial origin. 
We constrain PBH formation models within a hierarchical Bayesian inference framework based on deep learning techniques, finding best-fit values for distinctive features of these models, including the PBH initial mass function, the fraction of PBHs in dark matter, and the accretion efficiency.
The presence of several spinning binaries in the GWTC-2 dataset favors a scenario in which PBHs accrete and spin up.
Our results indicate that PBHs may comprise only a fraction smaller than $0.3 \%$ of the total dark matter, and that the predicted PBH abundance is still compatible with other constraints.
\end{abstract}

\maketitle

\section{Introduction}
Gravitational wave~(GW) astronomy is reshaping our understanding of the Universe. When taken individually, the black hole (BH) binary merger events announced before October 2020 by the LIGO-Virgo Collaboration~(LVC)~\cite{LIGOScientific:2018mvr,LIGOScientific:2020stg,Abbott:2020khf,Abbott:2020tfl} have firmly demonstrated that BHs form binaries that can merger within a Hubble time; that at least some of them have nonzero spins, possibly inducing precession in the orbital motion (see e.g.~\cite{LIGOScientific:2020stg}); and that they can exist in mass ranges that challenge the current stellar-formation paradigm~\cite{LIGOScientific:2020stg,Abbott:2020khf,Abbott:2020tfl}. Finally, their coalescence signal is fully consistent with the predictions of general relativity, providing novel and stringent bounds on modified gravity theories~\cite{Berti:2015itd,LIGOScientific:2019fpa,Abbott:2020jks}.

The recently released GWTC-2 dataset from the first part of the third observing run (O3a)~\cite{Abbott:2020niy} marks the onset of a transition from individual-event analyses to population studies: the number of BH merger events detected so far has more than quadrupled compared to the O1-O2 runs, making population studies particularly relevant~\cite{Abbott:2020gyp}.

Meanwhile, GW observations have renewed interest in trying to understand the nature of the observed population of BHs, from an astrophysical, cosmological, and theoretical standpoint~\cite{Barack:2018yly}. 

The two most popular astrophysical formation channels are isolated binary evolution in the field and dynamical formation in clusters (see e.g.~\cite{Mandel:2018hfr,Mapelli:2018uds} for reviews).  For isolated binaries, a common-envelope phase in between the formation of the two BHs is usually invoked to harden the binary and catalyze mergers. Alternatively, dynamical channels predict that binary BHs form and harden through three-body encounters in dense stellar clusters. Other popular binary BH formation scenarios include chemically homogenous evolution~\cite{Marchant:2016wow,deMink:2016vkw}, AGN disks~\cite{Leigh:2017wff,Stone:2016wzz,Bartos:2016dgn}, and secular interactions in triple systems~\cite{Silsbee:2016djf,Hoang:2017fvh,Fragione:2018yrb}. Different formation pathways leave different imprints on the properties of the binary BH population, including the binary masses, spins, eccentricities, and redshift evolution. Measuring these distributions informs us on the environment in which binary BHs form and evolve~\cite{Zevin:2017evb,Taylor:2018iat,Wysocki:2018mpo, Roulet:2018jbe,LIGOScientific:2018jsj}.

In this paper we will focus on another interesting possibility: that some of (if not all) the detected BH binaries are of primordial origin, i.e. they are composed of primordial black holes~(PBHs), whose formation took place in the early epochs of the Universe (see Refs.~\cite{Sasaki:2018dmp,Carr:2020gox, Green:2020jor} for recent reviews). This scenario is  also motivated by the fact that PBHs could comprise the entirety (or a fraction) of the dark matter~(DM) in the Universe, and several studies on the confrontation between the PBH scenario and GW data have been performed so far in the literature~\cite{Sasaki:2016jop,Bird:2016dcv,Clesse:2016vqa,Wang:2016ana, Ali-Haimoud:2017rtz, Raidal:2018bbj, Hutsi:2019hlw, Vaskonen:2019jpv, Gow:2019pok,DeLuca:2020qqa, Jedamzik:2020ypm, Jedamzik:2020omx, Hall:2020daa, DeLuca:2020sae, DeLuca:2020jug}.

Our aim is to constrain PBH population models using the entire dataset of BH binaries reported in the GWTC-2. We shall do so by applying hierarchical Bayesian inference based on deep learning techniques to find the best-fit parameters of the PBH model (including the PBH abundance), which can be then confronted to constraints coming from other observations.

The outline of this paper is as follows.
In Sec.~\ref{sec:Simulation} we describe the PBH simulations used in this study.
In Sec.~\ref{sec:Method} we review the data analysis pipeline, and
in Sec.~\ref{sec:Results} we apply it to public LVC GWTC-2 data.
In Sec.~\ref{sec:Constraints} we compare GW constraints on the fraction of PBHs in DM with those coming from other observations.
In Sec.~\ref{sec:Discussion} we summarize our findings and highlight future research directions.
Throughout this paper we use geometrical units ($G=c=1$).

\section{Key Predictions of the PBH scenario }
\label{sec:Simulation}

In this section we review the main theoretical predictions for the PBH scenario. More technical details can be found in Ref.~\cite{DeLuca:2020qqa} and references therein. 
After reviewing PBH formation, we report the key steps necessary to compute the merger rate, depending on the PBH mass function and abundance. We also summarize our model for accretion onto PBHs in binaries, because accretion has been shown to affect in a critical way the mass ratio, spins, merger rate, and overall abundance of the PBH population~\cite{DeLuca:2020bjf,DeLuca:2020fpg,DeLuca:2020qqa}.

There are several models giving rise to a cosmologically significant population of PBHs. In the most likely scenario, the formation of PBHs occurs from the collapse of large overdensities in the primordial Universe, when radiation dominates the energy density budget~\cite{Blinnikov:2016bxu,Ivanov:1994pa,Ivanov:1997ia}. 
The formation of a PBH of mass $m$ takes place deep in the radiation-dominated era at a typical redshift $z_\ii \simeq 2 \cdot 10^{11} (m/M_\odot)^{-1/2}$.
The resulting mass distribution depends on the characteristic size and statistical properties of the density perturbations, directly connected to the curvature perturbations produced during the inflationary epoch. A useful model-independent parametrization of the mass function at the formation redshift $z_\ii$ (here and below we will use a subscript ``i''  to indicate quantities evaluated at the formation epoch) is represented by the log-normal function
\be
\psi(m,z_\ii) = \frac{1}{\sqrt{2\pi}\sigma  m} \exp \lp -\frac{\log ^2 (m/M_c)}{2 \sigma^2} \rp
\ee
in terms of its width $\sigma$ and reference mass scale $M_c$. Such a mass function describes 
a population arising from a symmetric peak in the power spectrum of curvature perturbations and captures a wide variety of models~\cite{Dolgov:1992pu,Carr:2017jsz}. 

Since large perturbations tend to have nearly-spherical shape~\cite{bbks} and the collapse takes place in a radiation-dominated Universe, the initial adimensional spin parameter $\chi \equiv |J|/m^2$ is expected to be below the percent level~\cite{DeLuca:2019buf,Mirbabayi:2019uph}, with the characteristic value given by
\be
\chi_{\text{\tiny i}}
\sim 10^{-2} \sqrt{1-\gamma^2},
\ee
in terms of 
the width parameter $\gamma$ of the power spectrum~\cite{DeLuca:2019buf}.
 
During the cosmological history, PBHs in binaries may undergo a period of significant baryonic mass accretion, which impacts their individual masses~\cite{Ricotti:2007jk,Ricotti:2007au,zhang} and spins~\cite{bv,DeLuca:2020bjf,DeLuca:2020qqa}.
As the typical size of a binary is smaller than its corresponding Bondi radius, the infall of gas is driven by the binary system as a whole. This means that both PBHs experience accretion from the gas with an enhanced density. Accretion onto the two individual  PBHs is also modulated by their masses  and orbital velocities.
 By defining the mass ratio as $q \equiv m_2/m_1 \leq 1$, one can write the individual accretion rates as
\begin{align}
\dot m_1 = \dot m_\text{\tiny bin}  \frac{1}{\sqrt{2 (1+q)}}, 
\qquad 
\dot m_2 = \dot m_\text{\tiny bin}  \frac{\sqrt{q} }{\sqrt{2 (1+q)}},\label{M1M2dotFIN}
\end{align}
in terms of the Bondi-Hoyle mass accretion rate of the binary system
\be \label{R1bin}
\dot m_\text{\tiny bin} = 4 \pi \lambda m_H n_\text{\tiny gas} v^{-3}_\text{\tiny eff} M^2_\text{\tiny tot},
\ee
where $M_\text{\tiny tot} = m_1 + m_2$ is the total mass and dots denote derivatives with respect to time.
The expression above is explicitly dependent on the binary's effective velocity $v_\text{\tiny eff}$ relative to the baryons with cosmic mean density $n_\text{\tiny gas}$ and hydrogen mass $m_H$.
The accretion formula~\eqref{R1bin} adopts the Newtonian approximation; as recently pointed out in Ref.~\cite{Cruz-Osorio:2020dja}, general-relativistic effects may lead to a significant increase in the mass accretion rate.  However, since the accretion rate decreases by several orders of magnitude for PBH masses $\lesssim 10 M_\odot$~\cite{Ricotti:2007au},  even an order-of-magnitude increase in $\dot m$ does not change the predictions of our model significantly.
The accretion parameter $\lambda$ tracks the effects of the Hubble expansion, the coupling of the CMB radiation to the gas through Compton scattering, and the gas viscosity~\cite{Ricotti:2007jk}. 
Also, since PBHs in the mass
range of interest for LVC observations can comprise only a fraction of the DM due to the current constraints on their abundance, accretion onto PBHs should also include the presence of an additional DM halo forming around the PBHs~\cite{Ricotti:2007au,Adamek:2019gns,Mack:2006gz} (either isolated or in binaries). The DM halo acts as a catalyst enhancing the gas accretion rate, and its effect is taken into account in $\lambda$ (see Appendix~B of Ref.~\cite{DeLuca:2020bjf} for details). 
We account for the sharp decrease in the accretion efficiency around the epoch of structure formation~\cite{Hasinger:2020ptw,raidalsm,Ali-Haimoud:2017rtz} by defining a cutoff redshift $z_\text{\tiny cutoff}$ after which we neglect accretion. 
Because of the uncertainties in the model (such as X-ray pre-heating~\cite{Oh:2003pm}, details of the structure formation, and feedbacks of local, global~\cite{Ricotti:2007au,Ali-Haimoud:2016mbv} and mechanical type~\cite{Bosch-Ramon:2020pcz}), the cutoff redshift is relatively unconstrained. At variance with previous work~\cite{bv,DeLuca:2020bjf,DeLuca:2020qqa}, in which $z_\text{\tiny cutoff}$ was fixed to some reference value, here we consider it as a parameter of the PBH accretion model (more precisely, a ``hyperparameter'' of our model, see below) that we will fit to GW data, along with the other model parameters.

A consequence of PBH accretion in binaries is that the lighter binary component always accretes more efficiently, and therefore an initial mass ratio different from unity would grow as 
\be
\dot q = q \lp \frac{\dot m_2}{m_2} -  \frac{\dot m_1}{m_1} \rp >0.
\ee
Accretion also modifies the PBH mass distribution, making it broader at high masses and producing a high-mass tail that can be orders of magnitude above its corresponding value at formation. Furthermore, accretion also affects the total fraction of PBHs in DM $f_\PBH$ in a redshift-dependent fashion according to the relation \cite{DeLuca:2020fpg} \be\label{fPBHevo}
f_\PBH (z) = \frac{\langle m (z) \rangle}{\langle m (z_\ii) \rangle (f^{-1}_\PBH (z_\ii) - 1 ) + \langle m (z) \rangle},
\ee
where
\be
\langle m (z) \rangle = \int \d m \, m \, \psi(m, z)
\ee
is the average mass. This effect is of crucial importance when comparing the physical parameters to the existing experimental constraints~\cite{DeLuca:2020fpg}, as we will do below.

Since the fraction of PBHs which underwent mergers is of the order of ${\cal O}( 10^{-2} f_\PBH^{16/37})$~\cite{Liu:2019rnx,Wu:2020drm}, the overall PBH population is largely dominated by isolated PBHs, for which accretion is described in detail in Refs.~\cite{DeLuca:2020bjf,DeLuca:2020fpg} and references therein. This implies that the fraction of second-generation PBH mergers  is expected to be negligible in the LIGO/Virgo band~\cite{DeLuca:2020bjf,Liu:2019rnx,Wu:2020drm}, contrarily to the astrophysical scenario, in which second-generation mergers may play a significant role (see e.g.~\cite{Gerosa:2017kvu,Fishbach:2017dwv,Baibhav:2020xdf,Kimball:2020opk}).

In addition to changing the masses, the infalling accreting gas onto a PBH can carry angular momentum, which crucially determines the geometry of the accretion flow and the evolution of the PBH spin~\cite{Berti:2008af}. For accretion onto a PBH binary, the nonspherical geometry can give rise to a geometrically thin accretion disk as long as the accretion rate (normalized to the Eddington rate) is larger than unity~\cite{Ricotti:2007au,Shakura:1972te, NovikovThorne}. Only when this condition is satisfied the angular momentum transfer on each PBH is very efficient, and mass accretion is accompanied by an increase of the PBH spin according to the equation
\be
\dot \chi = g(\chi) \frac{\dot {m}}{m},
\ee
in terms of the function $g(\chi)$, which is derived using the geodesic model of disk accretion \cite{Bardeen:1972fi} (see e.g.~\cite{Brito:2014wla,volo,DeLuca:2020qqa}). We consider Thorne's spin limit $\chi_\text{\tiny max} = 0.998$, which is dictated by radiation effects~\cite{thorne} (see also~\cite{Gammie:2003qi}).

In particular, besides the angular-momentum transfer, the geometry of the disk crucially depends also on the accretion rate. If the mass accretion rate is sub-Eddington and nonspherical, an advection-dominated accretion flow~(ADAF) may form~\cite{Narayan:1994is}. For accretion rates close to the Eddington limit, one expects the formation of a geometrically thin disk~\cite{Shakura:1972te,Ricotti:2007au}, for which the angular momentum transfer can be described with a geodesic model~\cite{Bardeen:1972fi}. For moderately super-Eddington rates, the accretion luminosity might be strong enough that the disk ``puffs up'' and becomes slim~\cite{1988ApJ...332..646A}.
A larger accretion rate would imply geometrically thicker disks, with possible differences in the accretion luminosity and angular momentum transfer~\cite{1988ApJ...332..646A}. 
In such a scenario, accretion of angular momentum would be more complex, even though numerical simulations suggest that the time-scale for spin evolution is not significantly modified~\cite{Gammie:2003qi}. For the PBH masses under consideration, one does not exceed the Eddington accretion rate significantly, and therefore the thin-disk model should provide a reasonable approximation.

One of the clear predictions of the primordial scenario is that the spin of the lighter PBH in the binary is larger than that of the heavier PBH, because the spin-up is mainly produced by accretion and the secondary component typically accretes more.

A key observable inferred by GW observations is the binary's effective spin parameter
\begin{equation}
\label{chieff}
\chi_\text{\tiny eff} \equiv \frac{\chi_1 \cos{\theta_1} + q \chi_2 \cos{\theta_2}}{1+q}\,,
\end{equation}
defined in terms of the individual BH spin magnitudes $\chi_j$ ($j=1,2$), and the angles $\theta_1$ and $\theta_2$ between the orbital angular momentum and the individual spin directions. 
When forming PBH binaries, the spin directions are expected to be uncorrelated, and therefore uniformly distributed on the two-sphere.

When accretion is relevant in spinning up the PBHs, the accretion geometry is complex and the spin directions are expected to remain uncorrelated, as the orientations of the disks formed around each PBH are independent.
This is because the characteristic orbital distance is comparable to (or smaller than) the individual PBH Bondi radii. This scenario differs, for instance, from the common envelope phase giving rise to astrophysical BH binary mergers: in that case both of the individual BH spins are expected to be aligned with the orbital angular momentum of the binary (see e.g.~\cite{Gerosa:2013laa,Belczynski:2017gds,Gerosa:2018wbw,Steinle:2020xej}).

In the absence of primordial non-Gaussianities, the PBH locations in space at the formation epoch follow a Poisson distribution~\cite{Ali-Haimoud:2018dau,Desjacques:2018wuu,Ballesteros:2018swv,MoradinezhadDizgah:2019wjf}. Depending on the initial abundance  $f_\PBH(z_\ii)$ and mass function $\psi(m,z_\ii)$, one can compute how often binaries form in the early Universe by evaluating the probability that a binary system decouples from the Hubble flow. The initial distribution of the orbital parameters also depends on the spatial distribution of the surrounding population of PBHs, as well as density perturbations adding an  initial torque to the binary system. On top of that, following Ref.~\cite{Raidal:2018bbj}, one can introduce a correction $S$ to the merger rate, accounting for the possible disruption of binaries due to their interaction with the environment at high redshifts. Finally, introducing the effect of accretion, and following the discussion in Ref.~\cite{DeLuca:2020qqa} and references therein, one can compute the PBH merger rate as follows
\begin{align}
\label{diffaccrate}
& \d R
 = 
\frac{1.6 \times 10^6}{{\rm Gpc^3 \, yr}} 
f_\PBH^{\frac{53}{37}} (z_\ii)  
\eta^{-\frac{34}{37}}(z_\ii)
\lp \frac{t}{t_0} \rp^{-\frac{34}{37}}  
 \lp \frac{M^\ii_\text{\tiny tot}}{M_\odot} \rp^{-\frac{32}{37}}  \nonumber \\
& \times
S\lp M^\ii_\text{\tiny tot}, f_\PBH (z_\ii)  \rp
\psi(m^\ii_1, z_\ii) \psi (m^\ii_2, z_\ii) \nonumber \\
& \times 
 \exp\llp \frac{12}{37}\int_{t_\ii} ^{t_\text{\tiny cutoff}} \d t \lp \frac{\dot M_\text{\tiny tot}}{M_\text{\tiny tot}} + 2 \frac{\dot \mu}{\mu} \rp
\rrp  \nonumber \\
& \times \lp \frac{\eta (z_\text{\tiny cutoff})}{\eta (z_\text{\tiny i}) } \rp^{3/37} \lp \frac{ M_\text{\tiny tot}(z_\text{\tiny cutoff})}{M_\text{\tiny tot} (z_\text{\tiny i}) }
\rp^{9/37} \d m^\ii_1 \d m^\ii_2,
\end{align}
where 
$\mu= m_1 m_2/M_\text{\tiny tot}$, $\eta = \mu/M_\text{\tiny tot}$, and $t_0$ is the current age of the Universe.
The corrective factors in the last two lines account for the evolution of the masses from the initial redshift $z_\ii$ to the cutoff redshift $z_\text{\tiny cutoff}$ and the corresponding shrinking of the semimajor axis of the orbit due to accretion~\cite{DeLuca:2020qqa} (see also~\cite{Caputo:2020irr}). This effect dominates and drives the binary evolution up to the cutoff redshift, after which the binary evolution is uniquely driven by energy loss through GW emission~\cite{Peters:1963ux,Peters:1964zz}.
Notice  that the merger rate at small redshift ($z<z_\text{\tiny cutoff}$) has a universal scaling 
given by $t^{-34/37}$, which is independent of the other model parameters. Also,  for the small values of $f_\PBH$ which will be selected out in our analysis (see Sec.~\ref{sec:Results}), the clustering of PBHs plays no role~\cite{Inman:2019wvr,DeLuca:2020jug}.

\begin{table}
\caption{Event parameters ($\bm{\theta}$) of the binary and hyperparameters ($\bm{\lambda}$) of the PBH model used in this work.}
\begin{tabularx}{\columnwidth}{l X}
\hline
\hline
Event parameters $\bm{\theta}$ &  \\
$m_1$ & Source-frame primary mass \\
$m_2$ & Source-frame secondary mass \\
$\chi_{\rm eff}$ & Effective spin  \\
$z$ & Merger redshift \\
\hline
Hyperparameters $\bm{\lambda}$ &  \\
$M_c$ & Peak reference mass of the log-normal distribution\\
$\sigma$ & Variance of the log-normal mass distribution\\
$f_\PBH$ & Fraction of PBHs in DM at formation \\
$z_\co$ & Accretion cutoff redshift \\
\hline
\hline
\end{tabularx}
\label{Tb:parameters}
\end{table}

In summary, the parameters describing each PBH binary, along with the ``hyperparameters'' describing the PBH model,  are listed in Table~\ref{Tb:parameters}.
In order to perform the analysis discussed in the following section, we built a catalog  of simulations, producing the PBH merger population for the full set of hyperparameters $\bm{ \lambda}$. Each element in the catalog contains a population of $10^5$ merger events with the characteristics described above. In the region of $(M_c, \sigma)$ of interest for our analysis, a value of the cutoff redshift around $z_\text{\tiny cutoff}\sim 30$ or larger would correspond to a scenario where accretion is negligible for most of the PBH merger population.

\section{Data Analysis}
\label{sec:Method}

In this section we give a brief introduction to the data analysis methods used in this work.
We refer the reader to more comprehensive descriptions in the literature~\cite{2019PASA...36...10T,Vitale:2020aaz,2019MNRAS.484.4008G,Mandel:2018mve}.
In short, we use an implementation of hierarchical Bayesian inference based on deep learning techniques to constrain PBH formation models given LVC data.

Astrophysical models predict intrinsic properties of individual GW sources (such as the masses and spins of the binary components), while GW interferometers record a time series of the GW-induced strain in the detectors.
Before we can compare the model to the data, we must convert these time series into astrophysically meaningful quantities through a parameter estimation process~\cite{2019PASA...36...10T}.
The LVC's Gravitational Wave Open Science Center provides the output of this parameter estimation process as a collection of posteriors characterizing the expectation value and uncertainty on the properties of individual merger events, i.e. $p({\bm \theta}|{\bm d}_i)$, where ${\bm \theta}$ is a vector of source parameters, and ${\bm d}_i$ labels the time series of the $i$-th event in the catalog.

By Bayes' theorem, given some data $\bm{d}$, the posterior probability of the signal from an astrophysical source with parameters $\bm{\theta}$ is $p(\bm{\theta}|\bm{d}) \propto p(\bm{d}|\bm{\theta})p(\bm{\theta})$,
where $p(\bm{d}|\bm{\theta})$ is the likelihood of observing the data given our model of the astrophysical signal and detector,
and $p(\bm{\theta})$ is our assumed prior on the source parameters.
The prior encodes our previous knowledge of the underlying physics (e.g., mass should not be negative),
and plays an important role in interpreting the results~\cite{Pankow:2016udj,Vitale:2017cfs}.  Within the PBH scenario, the prior distributions of the binary parameters according to the model of Sec.~\ref{sec:Simulation} is different from the priors used by the LVC, and this can affect the inference of the individual source parameters~\cite{Bhagwat:2020bzh}. 
Crucially, our procedure does not rely on the priors because we reweigh them, and it is therefore valid also in the PBH scenario.

A hierarchical Bayesian analysis parametrizes the choice of priors by some vector of hyperparameters $\bm{\lambda}$,
whose posterior distribution 
can be inferred from the data
\begin{align}
p(\bm{\lambda}|\bm{d}) \propto p(\bm{\lambda})\int p(\bm{d}|\bm{\theta})p_{\rm pop}(\bm{\theta}|\bm{\lambda})\d \bm{\theta},
\label{eq:HBA}
\end{align}
where $p(\bm{d}|\bm{\theta})$ is the single-event likelihood, $p(\bm{\lambda})$ is now a prior on the hyperparameters, and $p_{\rm pop}(\bm{\theta}|\bm{\lambda})$ is the \textit{population likelihood}, equivalent to a prior parametrized by some hyperparameters. 
The parameters describing single events (e.g. masses, redshifts) are also referred to as {\it event parameters}, while the hyperparameters describing the entire sample (e.g. the fraction of DM in PBHs) can be also referred to as {\it population parameters}. 

An astrophysical population model characterized by population parameters $\bm \lambda$ will predict some distribution of event parameters $\bm \theta$ such that the differential rate $\frac{\rm d r}{{\rm d} {\bm \theta} }(\bm \lambda)$ is given by
\begin{equation}
\frac{\rm d r}{{\rm d} {\bm \theta} }(\bm \lambda) = R(\bm \lambda) \,p_{\rm pop}(\bm \theta|\bm \lambda)\,,
\label{population}
\end{equation}
where $\int p_{\rm pop}(\bm \theta|\bm \lambda) {\rm d}\bm \theta=1$,
and the total rate $R(\bm \lambda)$ is typically measured in yr$^{-1}$.
The predicted number of events is $N(\bm \lambda)= \int \frac{\rm d r}{{\rm d} {\bm \theta} }(\bm \lambda) \d {\bm \theta} \times T_{\rm obs}$,
where $T_{\rm obs}$ is the duration of the observing run(s).
We can take into account selection effects caused by the sensitivity of the detectors through a function $0 \leq p_{\textrm{det}}(\bm \theta)\leq 1$, corresponding to the probability that an event with parameters $\bm \theta$ would be detectable. The observable distribution is then given by
\begin{equation}
\frac{\rm d r_{\rm det}}{{\rm d} {\bm \theta} }(\bm \lambda) = R(\bm \lambda) \,p_{\rm pop}(\bm \theta|\bm \lambda)\, p_{\rm det}(\bm \theta)\,,
\label{population2}
\end{equation}
and the expected number of observations is $N_{\rm det}(\bm \lambda)= \int \d {\bm \theta} 
(\rm d r_{\rm det}/{\rm d} {\bm \theta})(\bm \lambda) \times T_{\rm obs}$.

All of these ingredients enter the population posterior, which has the standard expression for an inhomogeneous Poisson process (cf.~\citep{2004AIPC..735..195L,2018PhRvD..98h3017T,2019MNRAS.486.1086M,2019PASA...36...10T} for detailed derivations). In particular, the population posterior reads
\begin{align}
p({\bm \lambda}|{\bm d}) &\propto \ \pi({\bm \lambda})  \,e^{- N_{\rm det} ({\bm \lambda})} N({\bm \lambda})^{N_{\rm obs}} \nonumber \\ &\times\prod_{i=1}^{N_{\rm obs}} \! \int \!\frac{p({\bm \theta_i}|{\bm d})}{\pi({\bm \theta_i})} p_{\rm pop}({\bm \theta_i}|{\bm \lambda}){\rm d}{\bm \theta_i} \,,
\label{eq:posterior}
\end{align}
where $N_{\rm obs}$ is the number of observations and $\pi(\bm \lambda)$ is some assumed population prior. If one wishes to exclude rate information from the inference, marginalizing over $N(\bm \lambda)$ with prior $\propto 1/N(\bm \lambda)$ yields~\cite{2018ApJ...863L..41F}
\begin{equation}
p({\bm \lambda}|{\bm d}) \!\propto\! \pi({\bm \lambda})  \prod_{i=1}^{N_{\rm obs}} \! \int \!\frac{p({\bm \theta_i}|{\bm d})}{\pi({\bm \theta_i})}
\frac{p_{\rm pop}({\bm \theta_i}|{\bm \lambda})}{\alpha(\bm \lambda)}
{\rm d}{\bm \theta_i} \,,
\label{eq:posteriormarg}
\end{equation}
where $\alpha(\bm \lambda)$ is the fraction of events one would detect given a population (also known as the selection bias), and
\begin{align}
\alpha(\bm \lambda) = \int p_{\rm pop}({\bm \theta'}|{\bm \lambda}) p_{\rm det}(\bm \theta') \d \bm \theta' = \frac{N_{\rm det}(\bm \lambda)}{N(\bm \lambda)}.
\label{eq:selectionFunction}
\end{align}
In order to accurately capture the detector response to account for the selection bias,
one should reweight the injection campaign released by LIGO~\cite{SelectionData} according to the method described in \cite{2019RNAAS...3...66F}.
However, the domain of masses in our model is larger than the domain where the injection campaign was performed ($2 M_\odot<m_1,m_2<100\ M_{\odot}$),
which means that the probability density function describing our model is not normalized within the domain of the injection campaign.
This induces severe inaccuracy in estimating the selection bias with the reweighting method, so we use a more crude but more well-behaved method to estimate the selection bias. 
For O1 and O2 data, it has been shown that approximating $p_{\textrm{det}}({\bm \theta})$ with the single-detector semianalytic approximation of Refs.~\cite{1993PhRvD..47.2198F,1996PhRvD..53.2878F} and a signal-to-noise ratio threshold of 8 yields results in good agreement with large-scale injection campaigns~\cite{2016ApJ...833L...1A,2018arXiv181112940T}.
We used \texttt{aLIGOEarlyHighSensitivityP1200087} (\texttt{aLIGOMidHighSensitivityP1200087}), as implemented in \texttt{pycbc}, as our O1-O2 (O3a) sensitivity curves.
To evaluate the selection bias integral of Eq.~\eqref{eq:selectionFunction} we can again use importance sampling.
Given a synthetic catalog, we compute $\alpha({\bm \lambda})$ by taking the average $p_{\rm det}({\bm \theta})$ of events in the catalog according to Eq.~\eqref{eq:selectionFunction}.
The network can be used both to evaluate the likelihood function and to generate new simulations.
At any given point in the hyperparameter space, we generate $10^5$ samples using our network, then evaluate the selection bias with the code described in Ref.~\cite{Wong:2020wvd}.

The event posterior probability distribution function $p(\bm{\theta}_i|\bm{d})$ is often given in the form of ${\cal S}_i$ discrete samples by a parameter estimation process~\cite{Veitch:2014wba,Ashton:2018jfp}.
As clear from Eq.~\eqref{eq:posteriormarg}, we weighted out
the priors on the event parameters $\pi({\bm \theta_i})$, so they do not contribute to $p({\bm \lambda}|{\bm d})$.
We can now make use of the posterior samples in the population inference,
thus avoiding unnecessary reevaluations of $p(\bm{d}|\bm{\theta})$ and significantly reducing the computation load for each population inference run.
The integral in Eq.~\eqref{eq:posterior} can be evaluated by using importance sampling, i.e. by computing the expectation value of  the prior-reweighted population likelihood,
which can be turned into a discrete sum over the samples of the event posterior probability distribution function
\begin{align}
p(\bm{\lambda}|\bm{d}) = \pi(\bm{\lambda})e^{- N_{\rm det} ({\bm \lambda})} N({\bm \lambda})^{N_{\rm obs}}\prod_{i=1}^{N_{\rm obs}}\frac{1}{{\cal S}_i}\sum_{j=1}^{{\cal S}_i} \frac{p_{\rm pop}(^j\bm{\theta}_i|\bm{\lambda})}{\pi(^j\bm{\theta}_i)},
\label{eq:populationPosterior_discrete}
\end{align}
where $j$ labels the $j$-th sample of the $i$-th event.

We train a deep learning emulator on the simulations described in Sec.~\ref{sec:Simulation} to evaluate the population likelihood $p_{\rm pop}(\bm{\theta}|\bm{\lambda})$.
Here we give some details on the network's architecture.
A more detailed discussion of the neural network and of the training procedure can be found in Refs.~\cite{Wong:2019uni,Wong:2020jdt}.
We use a masked autoregressive flow network~\cite{2017arXiv170507057P} with 10 hidden layers, each layer having 1024 units as our main architecture.
We train two variants using the same architecture and data, one with 
4 observables $\{m_1,q,\chi_\text{\tiny eff}, z\}$ and 3 hyperparameters $\{M_c,\sigma,z_\text{\tiny cutoff}\}$, and another one where we exclude $\chi_\text{\tiny eff}$ from the observables.
Note that we follow the LVC convention to enforce $m_1 > m_2$.
We do not train on $f_\text{\tiny PBH}$, because it affects only the overall rate, but not the shape of the population likelihood.
Our training set contains $2100$ simulations with different combinations of hyperparameters, and $10^5$ sample points in the observable space per simulation, adding up to $2.1\times 10^8$ sample points on the parameter-hyperparameter space in total.
As customary in training neural networks, $80\%$ of the data are used for training, $10\%$ are used for validation during the training to avoid overfitting, and the remaining $10\%$ is used to test the results independently.
We train the network for 100 epochs on a Nvidia K80 GPU to ensure convergence.
The code for the neural network is written in {\sc python} with {\sc pytorch}~\cite{paszke2017automatic}.

The final piece we need is the predicted number of events $N({\bm \lambda})$. In order to compute the latter, we first need to have a rate function, which requires running a full simulation, then summing over the rate for each individual event. Therefore, computing the number of events is as expensive as generating the simulation itself.  We employ a simple network to interpolate the rate as a function of the 4 hyperparameters, as described in Ref.~\cite{Wong:2020wvd}.  We have checked that the median error in our interpolation method is $\sim 0.04\%$, with $98\%$ of the errors being smaller than $3\%$, therefore the interpolation error is negligible compared to statistical uncertainties. Once we know the intrinsic merger rate $R(\bm \lambda)$, we can trivially obtain the observed merger rate as $R_{\rm det}(\bm \lambda) =\alpha(\bm \lambda) R(\bm \lambda)$ (see Eq.~\eqref{eq:selectionFunction}).
Once we have the rate function, we can multiply the rate by the effective observational time to obtain the predicted number of events.
For O1-O2 (O3a), there are $\sim 166.6$ $(183.3)$ days of coincident data.

Among all the binary events included in the GWTC-2 catalog, we use the same subset selected for the population analysis in Ref.~\cite{Abbott:2020gyp}. In particular, we exclude events with large false-alarm rate (GW190426, GW190719, GW190909) and events where the secondary binary component has mass smaller than $3 M_\odot$ (GW170817, GW190425, GW190814). Therefore, our dataset includes 44 events.

\begin{figure*}
\includegraphics[width=0.7\textwidth]{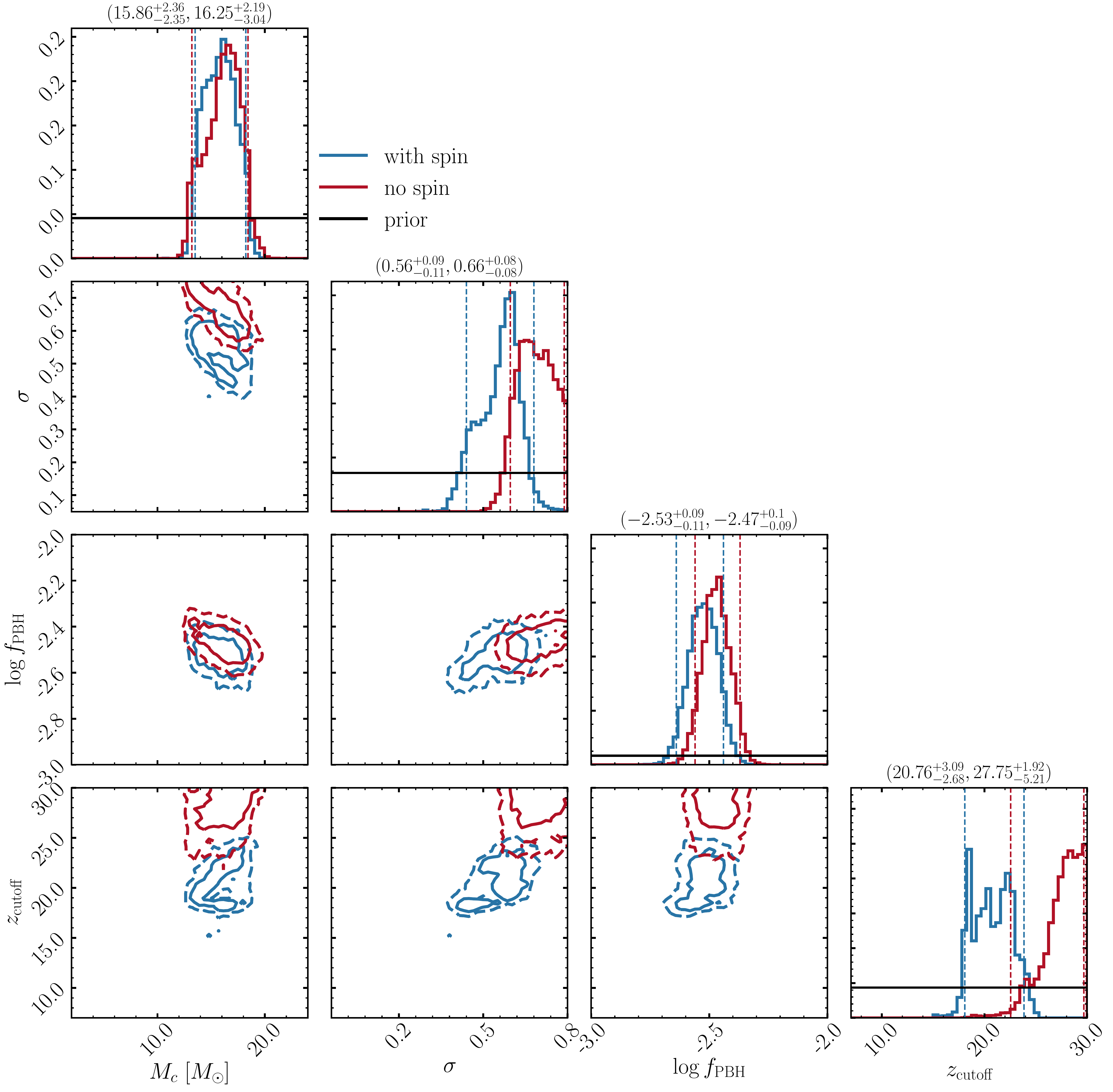}
\caption{Population posterior using the 44 GWTC-2 BH binary events. Blue lines are obtained using four observables $(m_1,\,m_2,\,\chi_{\rm eff},\,z)$, whereas red lines do not consider the effective spin in the inference.
Solid (dashed) contours represent $68\%$ ($95\%$) confidence intervals. The solid black lines indicate the priors assumed for the population hyperparameters. The first (second) number in parentheses is the hyperparameter range inferred by including (omitting) $\chi_{\rm eff}$ from the inference.}
\label{fig:Phenom_inference_nospin_O3}
\end{figure*}

Among the events with $m_2<3 M_\odot$, an electromagnetic counterpart was detected and identified with a kilonova only for GW170817~\cite{GBM:2017lvd}, showing that at least one of the binary components (and most likely both) is a neutron star. However, in the absence of a counterpart, it is much more uncertain to assess whether the light components of the other two events (GW190425 and GW190814) are indeed neutron stars. In fact it cannot be excluded that these events have a different, and possibly primordial, origin~\cite{Abbott:2020khf}. Here we excluded these events (identified by the LVC as neutron-star or mixed binaries) to conform with the LVC population analysis~\cite{Abbott:2020gyp}. Including them would be an interesting extension of our work. However, we expect that the inclusion of only two not particularly loud events (GW190425 and GW190814) out of $44+2$ potential binary BHs would not change our results significantly.

We adopt the {\sc Overall\_posterior} provided in \cite{PEresult_GWTC1} for events in GWTC-1, and the {\sc PublicationSamples} provided in \cite{PEresult_GWTC2} for events in the GWTC-2 catalog.
We apply both the model with effective spin and without effective spin to analyze the data, with a uniform population prior in the range where we trained our emulator.
We sample Eq.~\eqref{eq:populationPosterior_discrete} using the MCMC package \texttt{emcee}~\cite{2013PASP..125..306F}.

\section{Inference from the GWTC-2 catalog}\label{sec:Results}

In this section we describe the results of our analysis of GWTC-2 events.
The best-fit hyperparameters obtained are summarized in Table~\ref{Tb:results:parameters}.

\begin{table}
\caption{Hyperparameters of the PBH model inferred using $(m_1,m_2,\chi_{\rm eff},z)$ and GWTC-2 data.}
\vspace{0.3cm}
\begin{tabularx}{0.6\columnwidth}{X c}
\hline
\hline
$M_c [M_\odot]$ & $15.86 ^{+ 2.36}_{- 2.35}$   \\
$\sigma$ & $0.56^{+0.09}_{-0.11}$  \\
$\log f_\PBH$ & $-2.53^{+0.09}_{-0.11}$  \\
$z_\co$ & $20.76^{+ 3.09}_{-2.68}$  \\
\hline
\hline
\end{tabularx}
\vspace{-0.2cm}
\label{Tb:results:parameters}
\end{table}

Our analysis has several improvements with respect to the existing literature. First, many attempts to perform population inference neglected the role of accretion (see e.g.~\cite{Raidal:2018bbj,Wu:2020drm,Dolgov:2020xzo,Hall:2020daa}), which was shown to be relevant \cite{DeLuca:2020qqa}. We can only compare with studies of the GWTC-1 dataset which neglected accretion if we restrict to large values of $z_\co$. In this limit, our results are in general agreement with previous work. 
Recently, some of us~\cite{DeLuca:2020qqa} studied the impact of PBH accretion on the merger rate and on the distribution of binary parameters, inferring the PBH population properties from a maximum-likelihood analysis of the GWTC-1 dataset. For a fixed accretion $z_\co$, we have checked that those results are compatible with the present analysis. 
Here, for the first time, we treat $z_\co$ as a free hyperparameter and we infer its posterior distribution. 
A second major improvement with respect to Ref.~\cite{DeLuca:2020qqa} is that we now include the effective spin information in the Bayesian inference.

\begin{figure*}[ht!]
\includegraphics[width=0.4\textwidth]{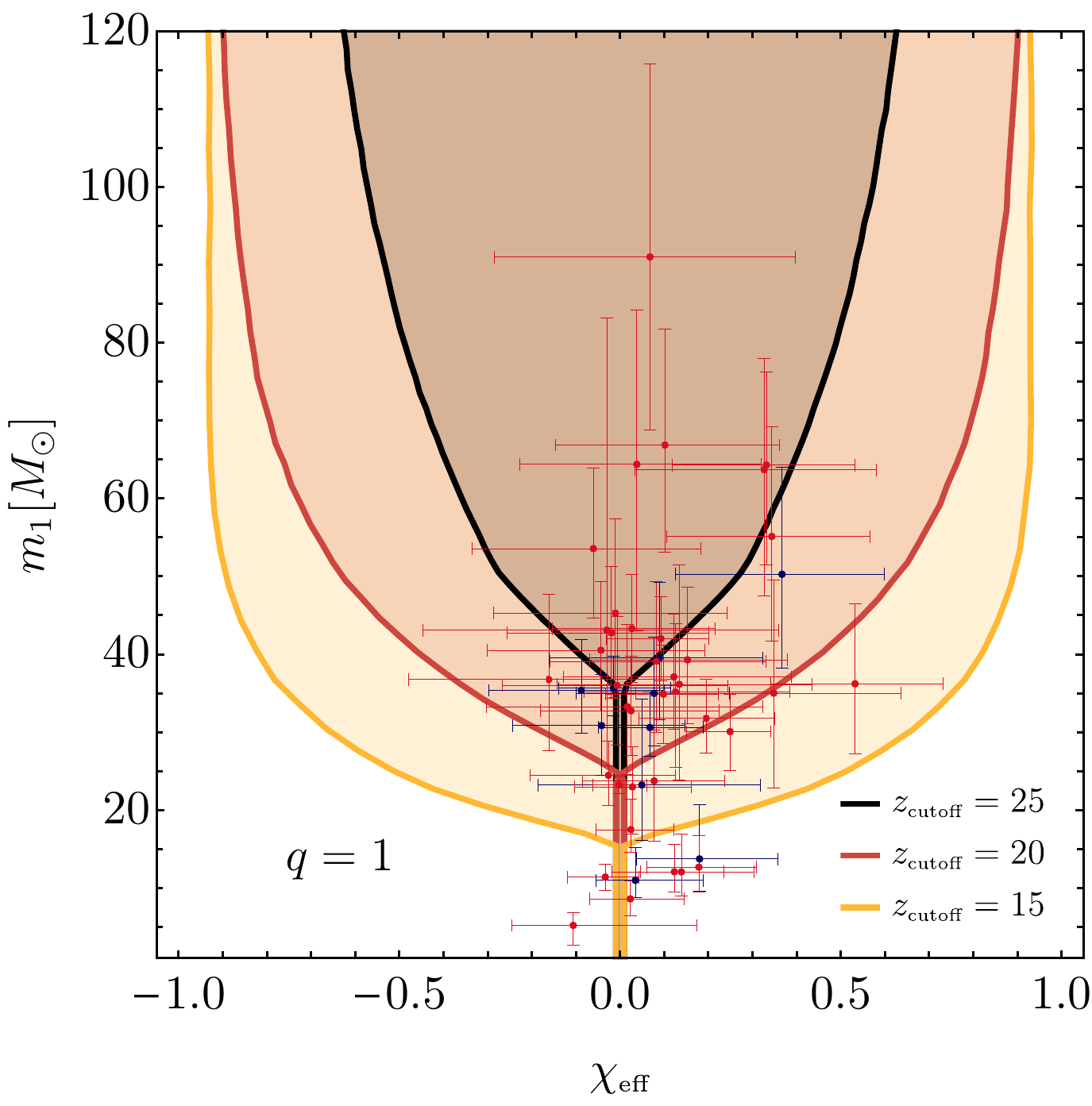}
\hspace{.4 cm}
\includegraphics[width=0.4\textwidth]{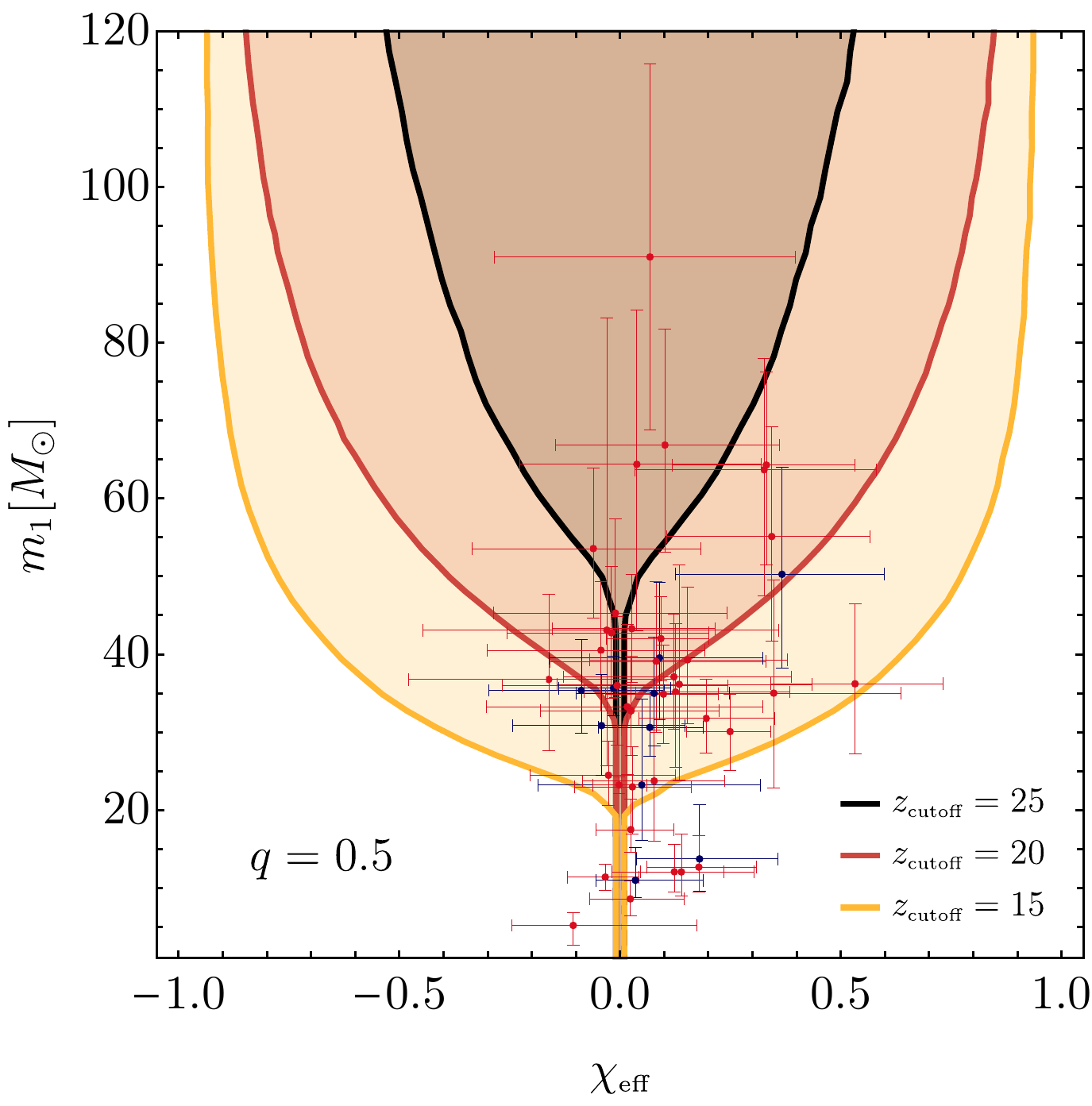}
\caption{Prediction for the $\chi_\text{\tiny eff}$ distribution as a function of the primary BH mass $m_1$ and cutoff redshift, at $2\sigma$ confidence level, for the best PBH scenario inferred from the GWTC-2 dataset. In blue we show the events from the GWTC-1; in red, the new events reported after the O3a observing run.} 
\label{fig:chieff_M1_O1O2O3a}
\end{figure*}

\begin{figure*}
 \includegraphics[width=0.44\textwidth]{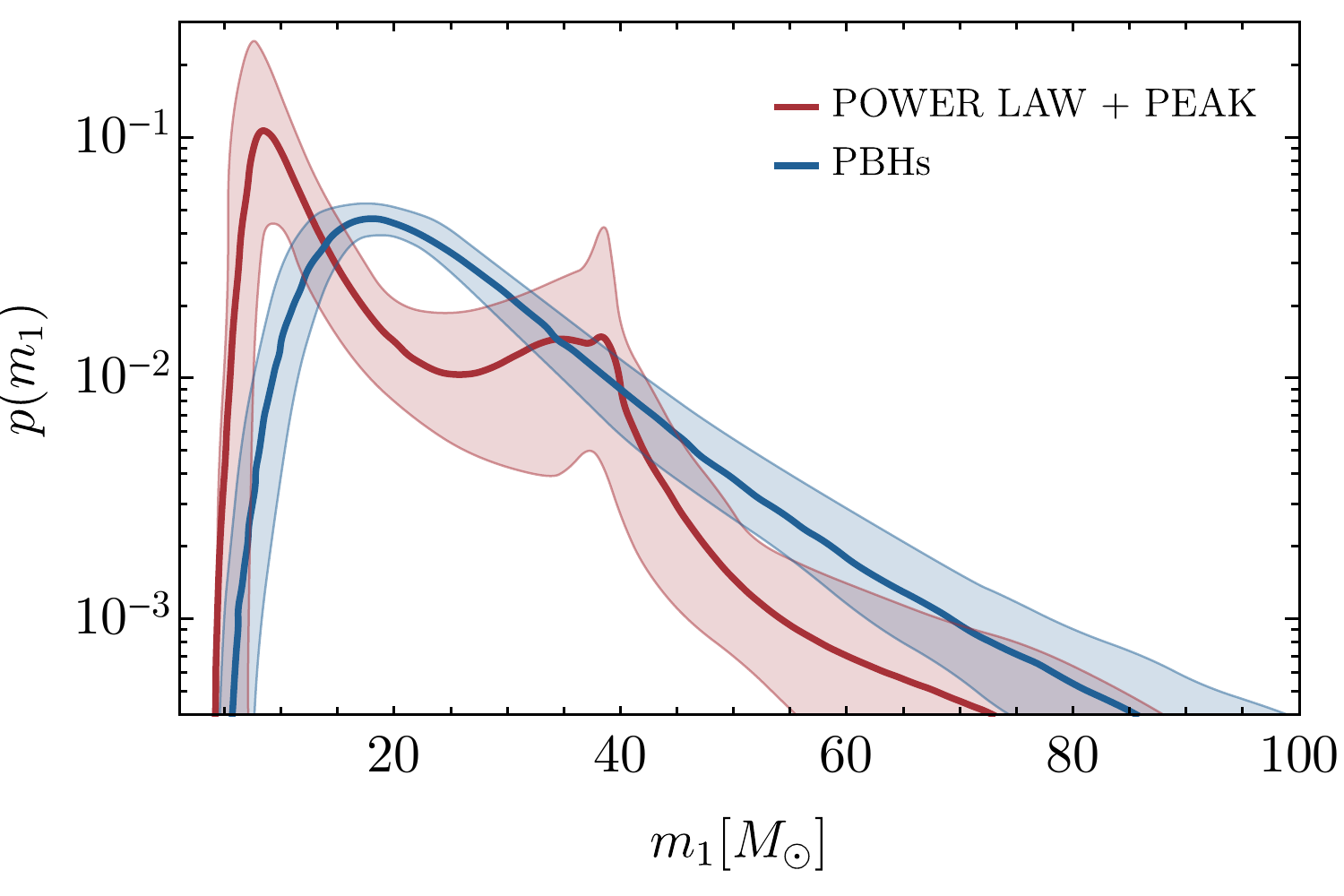}
 \hspace{.4 cm}
 \includegraphics[width=0.44\textwidth]{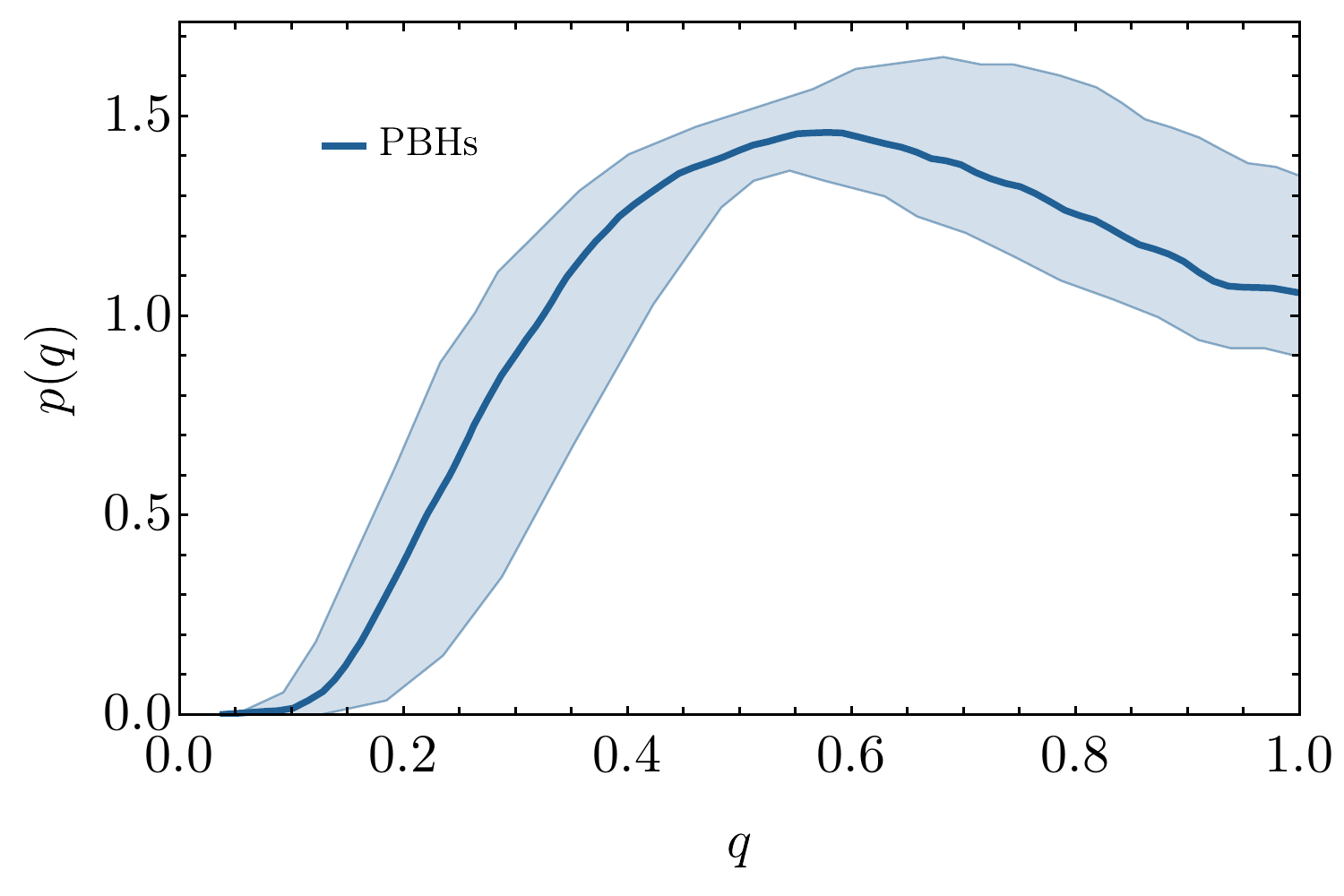}
 \includegraphics[width=0.44\textwidth]{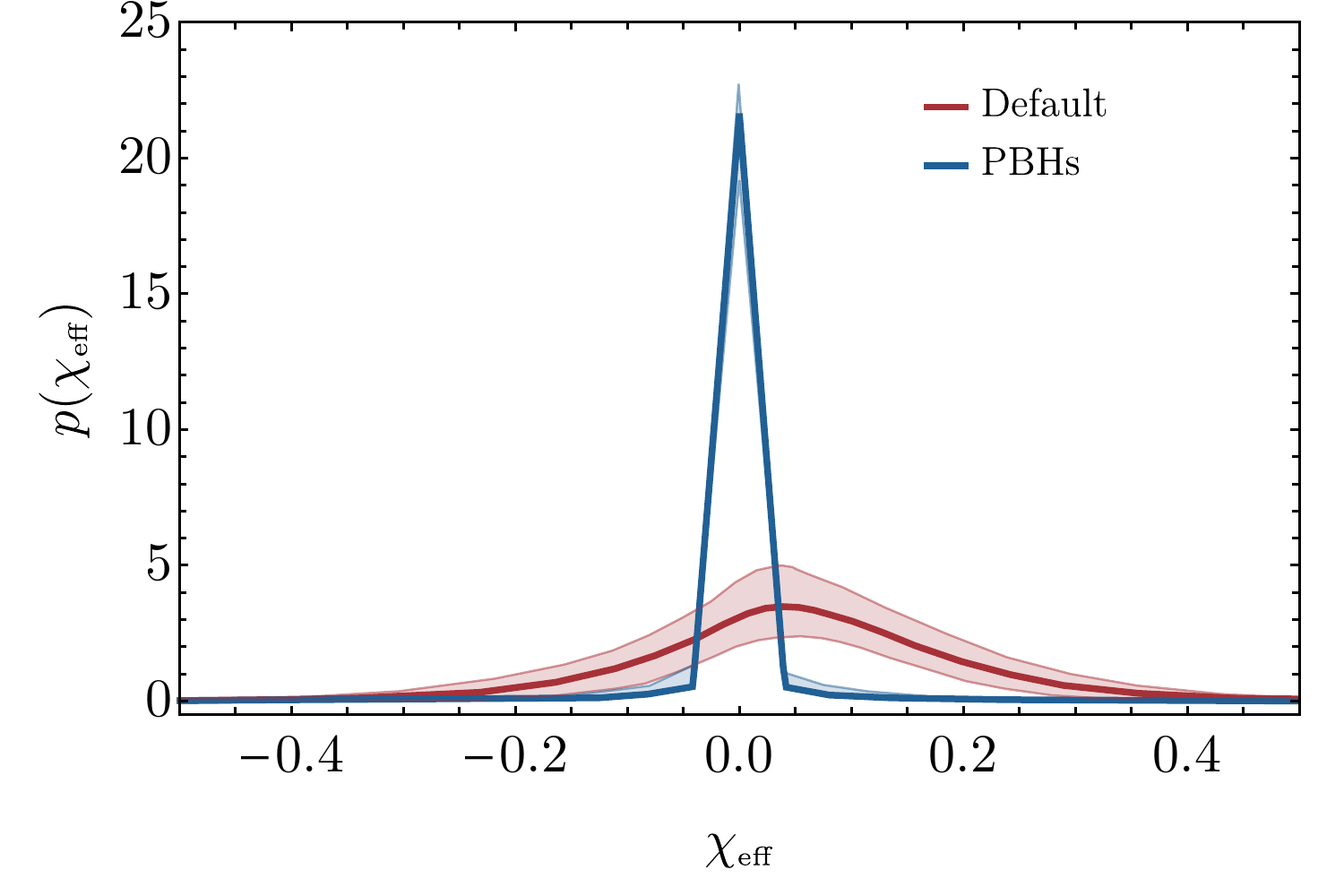}
 \hspace{.4 cm}
\includegraphics[width=0.44\textwidth]{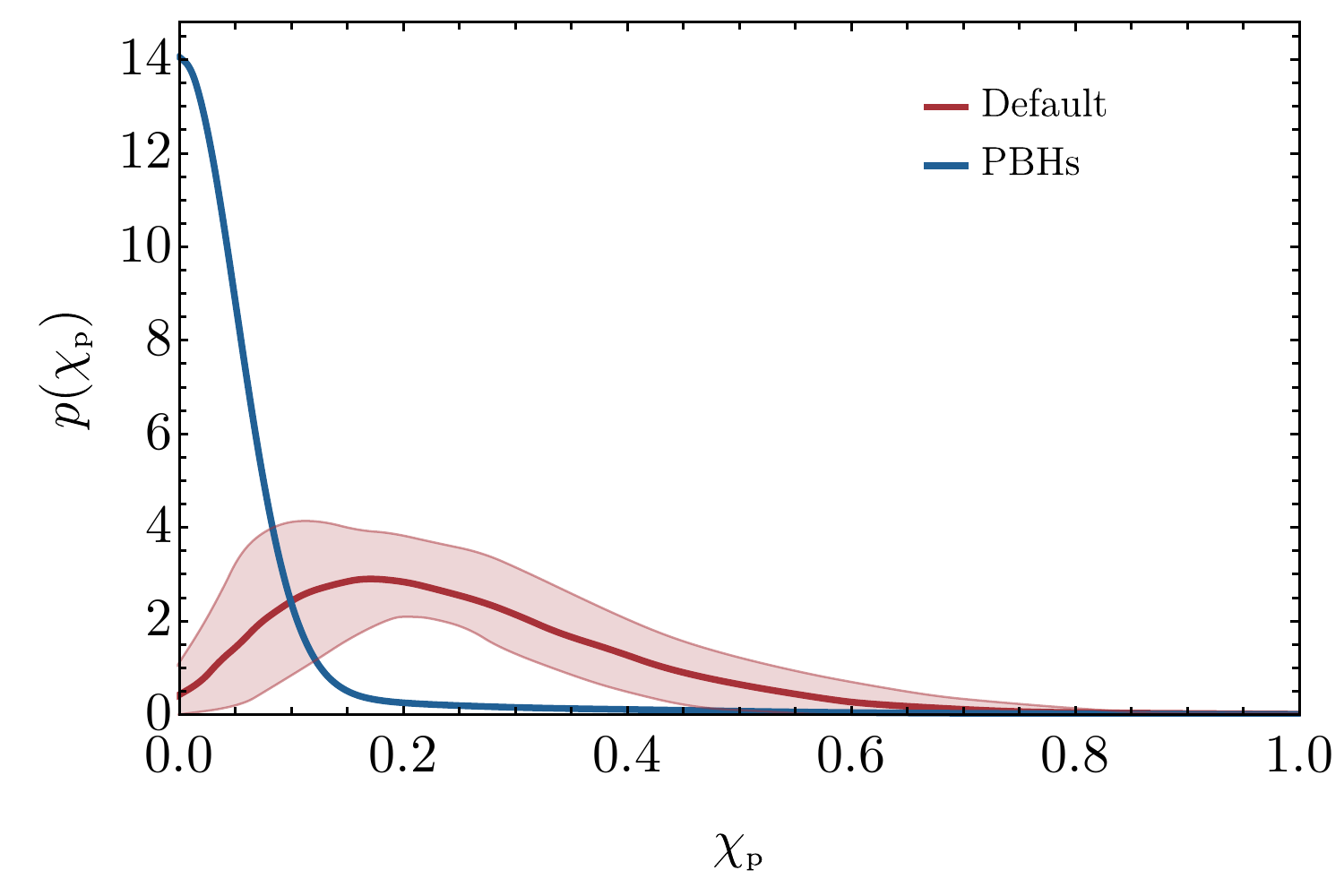}
\caption{Distribution of the primary mass $m_1$ (top left), mass ratio $q$ (top right), effective spin $\chi_\text{\tiny eff}$ (bottom left) and precession spin $\chi_\text{\tiny p}$ (bottom right) from our best-fit model when we include $\chi_\text{\tiny eff}$ (but not $\chi_\text{\tiny p}$) in the inference. For comparison, we also show the $90\%$ CL distributions found by the LVC in Ref.~\cite{Abbott:2020gyp} for astrophysical sources.}
\label{fig:populationdistpropertiesO1O2O3a}
\end{figure*}

In Fig.~\ref{fig:Phenom_inference_nospin_O3} we show the posterior distribution of PBH hyperparameters obtained by applying hierarchical Bayesian inference to the GWTC-2 dataset.
In order to highlight the information content coming from effective spin measurements, we compare inference results obtained with and without the use of spin information.
When we neglect $\chi_{\rm eff}$, the mass distribution of the events in the catalog favors high values of $z_\co$, making accretion less relevant.
When we include $\chi_{\rm eff}$, the best-fit value of the characteristic scale of the initial mass function $M_c$ decreases only slightly, but the posterior of $z_\co$ gets narrower and peaks at smaller values. This is because several events in O3a have effective spin not compatible with zero, and accretion is necessary to spin up PBHs. This also affects the posterior of $\sigma$ (the initial width of the PBH mass function), which gets narrower and peaks at a smaller value, because accretion broadens the mass function. As expected, the PBH abundance is found to be relatively stable with respect to changes of the other hyperparameters: we find $f_\text{\tiny PBH}\simeq 3 \cdot 10^{-3}$, indicating that this population of PBHs can comprise at most a subpercent fraction of the totality of the DM.

In Fig.~\ref{fig:chieff_M1_O1O2O3a} we show the 2$\sigma$ confidence intervals of the effective spin parameter distribution predicted by the PBH model in terms of the primary component mass, for different values of the mass ratio. The values of the cutoff redshift have been chosen around the 2$\sigma$ range obtained from best-fitting the primordial scenario considering the GWTC-2 dataset.
Following Refs.~\cite{DeLuca:2020bjf,DeLuca:2020qqa}, for each mass value $m_1$, we have averaged over the individual spin directions with respect to the total angular momentum assuming isotropic and independent distributions.
Because of the stronger impact of accretion on binaries with a larger total mass, one finds an enhancement of the  PBH spins with respect to the small values inherited at formation only above a certain  threshold.
The distributions shown in Fig.~\ref{fig:chieff_M1_O1O2O3a}
highlight the transition from initially vanishing values of the spins to large values depending on the binary masses and accretion strength.
Since the transition from negligible to large values of the spins is pushed towards smaller masses as the cutoff is reduced (i.e. stronger accretion),
the presence of several spinning binaries in the GWTC-2 catalog leads to a preference towards smaller values of $z_\co$.
Blue points with error bars are data from the GWTC-1, while red points are new detections from the O3a run, as reported by the LVC using agnostic priors.

In Fig.~\ref{fig:populationdistpropertiesO1O2O3a} we show the distribution of the most relevant binary parameters (primary mass $m_1$, mass ratio $q$, effective spin $\chi_{\rm eff}$, and precession spin $\chi_{\rm p}$) inferred from our best-fit model. On the top left, we plot the marginalized posterior probability for the primary mass $m_1$ for the PBH scenario, adding also for comparison the corresponding preferred result found in Ref.~\cite{Abbott:2020gyp} assuming a ``\textsc{Power-law + Peak}'' mass function. On the top right, we plot the marginalized distribution for the mass ratio in the PBH case. Notice that, due to the preferred relatively high value of the cutoff redshift and the significant width of the mass function, the distribution is peaked at $q \sim 0.5$ (had we found a smaller value of $z_\text{\tiny cutoff}$ the distribution would have peaked at higher values of the mass ratio as predicted by the PBH scenario with accretion~\cite{DeLuca:2020bjf,DeLuca:2020qqa}). On the bottom we plot the marginalized distributions for the effective spin parameter $\chi_\text{\tiny eff}$ and the precession spin $\chi_\text{\tiny p}$, which parametrizes the spin components perpendicular to the binary angular momentum responsible for the precession of the orbital plane, both for the PBH scenario and the so-called ``Default'' model, see Appendix D.1 of  Ref.~\cite{Abbott:2020gyp}. In both cases, the probability distributions inferred from the PBH model show a narrow peak around zero since the best-fit PBH mass function is dominated by 
relatively small masses, which are correlated with  small spins. This is not in contrast with the fact that we find a preference for an accreting PBH model due to the presence of several (moderately) spinning binaries in the catalog. Indeed, we stress that Fig.~\ref{fig:populationdistpropertiesO1O2O3a}
shows the population distribution, which does not account for selection effects (current detectors favor the observation of large masses).
This explains the difference with the ``Default'' model, for which masses and spins are not correlated, giving rise to a peak at nonvanishing spins and broader distributions.

\begin{figure}
\includegraphics[width=0.44\textwidth]{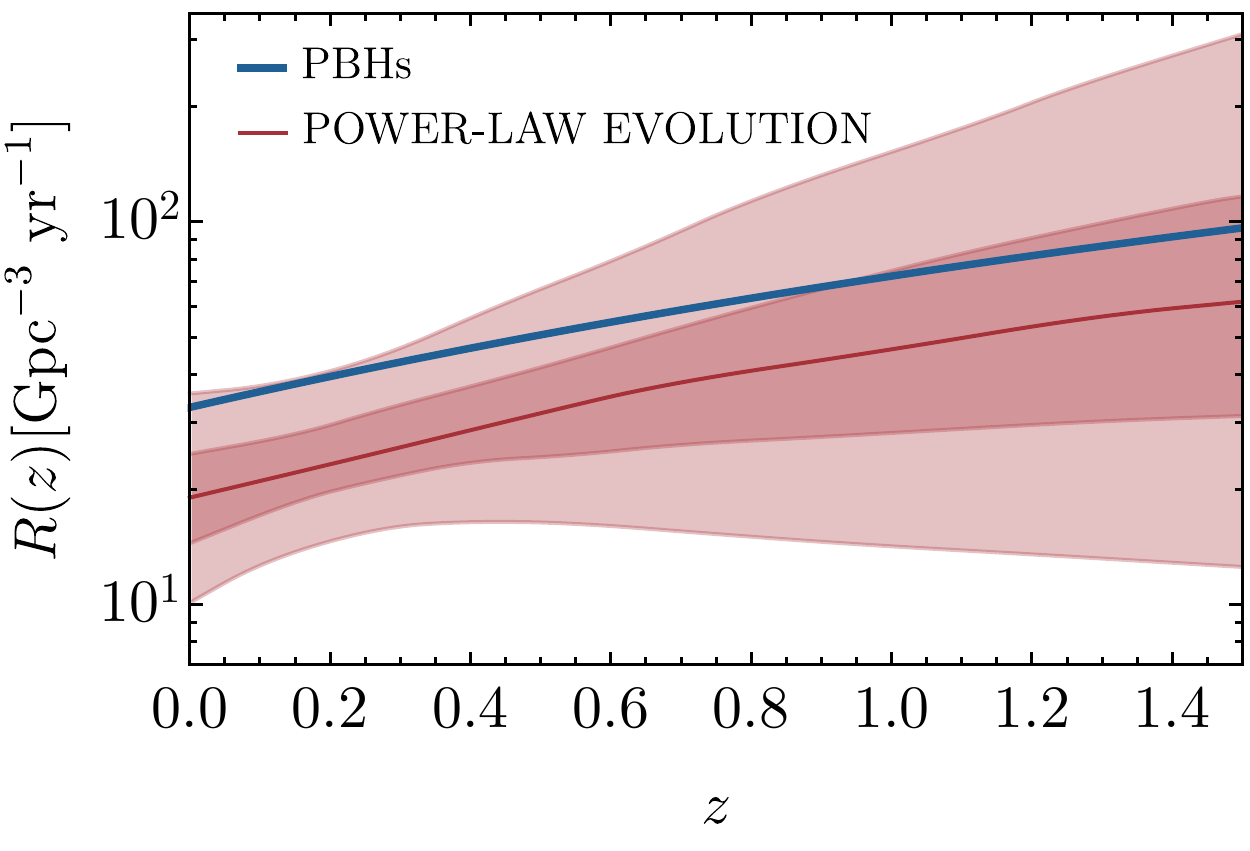}
\caption{
PBH merger rate density evolution given by Eq.~\eqref{diffaccrate} for the best-fit population (blue line). For comparison, we also show (in red) the $50\%$ ($90\%$) confidence level for the merger rate found by the LVC~\cite{Abbott:2020gyp} adopting a power-law evolution model for astrophysical sources.}
\label{fig:populationdistpropertiesO1O2O3a_rate}
\end{figure}

\begin{figure*}
\includegraphics[width=0.9\textwidth]{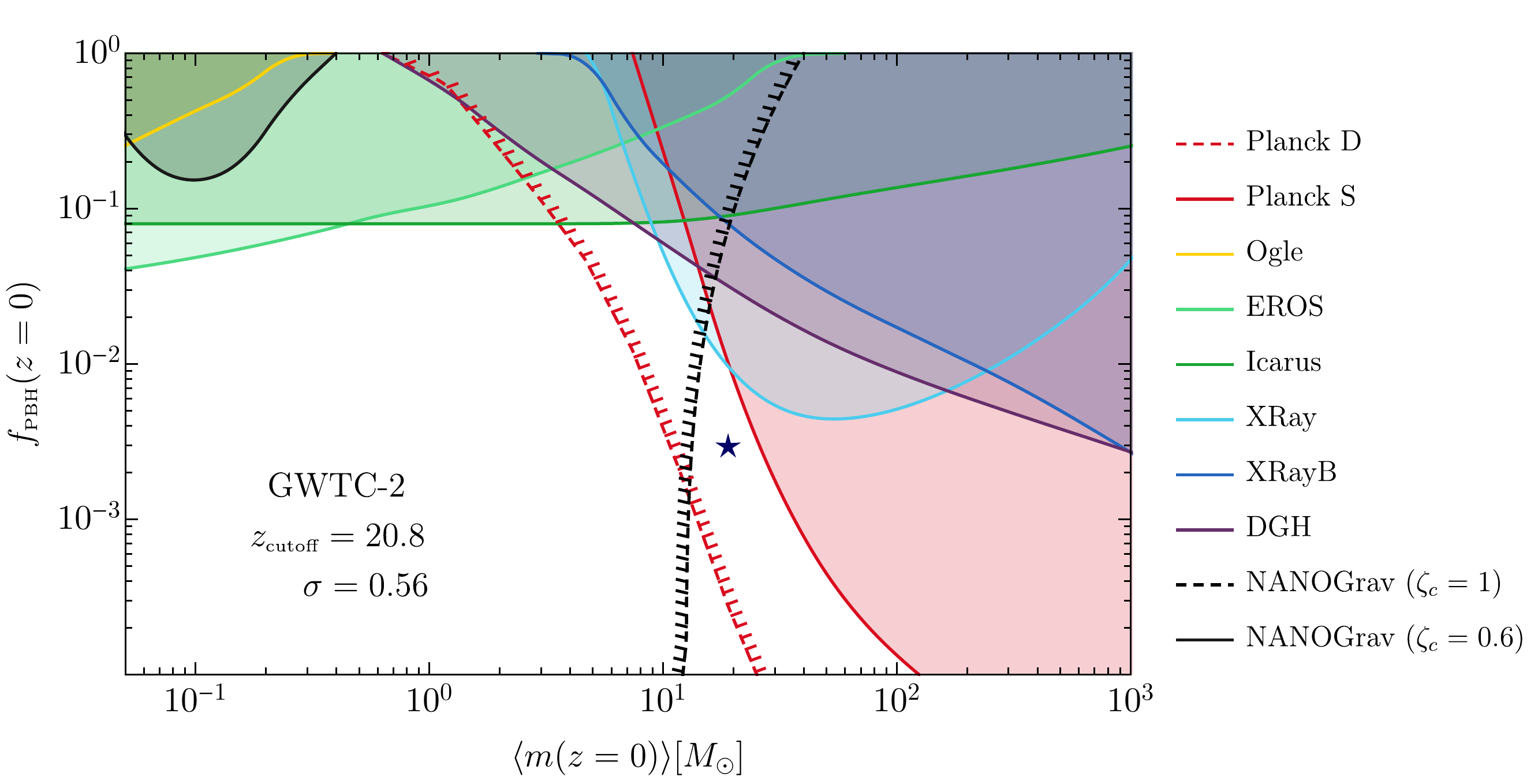}
\caption{
Summary of the constraints on the present PBH abundance $f_\PBH(z=0)$ as a function of the mean present PBH mass $\langle m(z=0) \rangle$. The blue star indicates the median values for the population parameters $\{ M_c, \sigma, f_\PBH, z_\text{\tiny cutoff} \}$ inferred with the GWTC-2 dataset.
We have indicated by the  red and black dashed lines, respectively, the bounds from Planck D and NANOGrav, which carry uncertainties (see the main text for details). The perpendicular dashes point towards the would-be excluded regions.}
\label{fig:constraintsO1O2O3a}
\end{figure*}

Finally, in Fig.~\ref{fig:populationdistpropertiesO1O2O3a_rate} we compare the prediction for the evolution as a function of redshift of the merger rate density $R(z)$, as given by Eq.~\eqref{diffaccrate} for the best-fit PBH model, with the power-law evolution model for astrophysical sources found in Ref.~\cite{Abbott:2020gyp}. Interestingly, the latter is slightly less steep than the $R(z)\propto (1+z)^{2.7}$ behavior predicted  from the star formation rate~\cite{Madau:2014bja}, although current observational errors are still too large to draw any conclusion. At any rate, the measured merger rate evolution is compatible with the PBH scenario, which predicts less mergers compared to the stellar-origin scenario in the high redshift side of the LVC horizon. 
Furthermore, the PBH scenario predicts a merger rate which increases monotonically also at redshifts beyond the LVC horizon,
while in the astrophysical case the rate is expected to decrease soon after the $z\sim 2$ star formation rate peak redshift (unless there is a significant contribution from Population III binaries, which could produce a second peak at large redshift). This difference is of particular interest for third-generation GW detectors, which may be able to detect mergers up to redshift $z \simeq {\cal O}(10^2)$~\cite{Sathyaprakash:2019yqt,Maggiore:2019uih}.

\section{Constraints on the PBH abundance}
\label{sec:Constraints}
In this section we compare the PBH abundance found to explain the observed BH merger events within the PBH scenario to other existing constraints.
We refer to the recent reviews~\cite{Sasaki:2018dmp,Carr:2020gox} and references therein for more details.

In the mass range of interest for our discussion, the most relevant constraints come from CMB anisotropies~\cite{Ali-Haimoud:2016mbv, Serpico:2020ehh}. PBHs start accreting gas in the early Universe in a process accompanied by emission of ionizing radiation, to which CMB temperature and polarization fluctuations are very sensitive. The constraints derived in \cite{Serpico:2020ehh} also take into account the catalysing effect of the early DM halo forming around  individual PBHs, as discussed in Sec.~\ref{sec:Simulation}. Due to uncertainties in the accretion physics, the authors analyze two alternative scenarios believed to bracket uncertainties: the accreting gas is either in a disk or in a spherical geometry (Planck D/S respectively).
The relevant electromagnetic emission takes place in the redshift range $300 \lesssim z \lesssim 600$. This means that the spherical model (Planck~S) is expected to be more accurate, as a thin accretion disk could form only at much smaller redshift~\cite{Ricotti:2007au}, as discussed in Sec.~\ref{sec:Simulation}. Also, since the relevant emission takes place at high redshift, its physics is independent of uncertainties in the accretion model due to the onset of structure formation.  Finally, as inferred from the results of the N-body simulation performed in Ref.~\cite{Inman:2019wvr}, clustering at that early epochs is not relevant (see also \cite{Hutsi:2019hlw}).

Other constraints come from comparing the {\it late time} emission of
electromagnetic signals from interstellar gas accretion onto PBHs with observations of galactic radio and X-ray isolated sources~(XRay)~\cite{Gaggero:2016dpq,Manshanden:2018tze} and 
X-ray binaries~(XRayB)~\cite{Inoue:2017csr}, Dwarf Galaxy Heating~(DGH) due to interactions of PBHs with the interstellar medium using data from Leo~T dwarf galaxy observations~\cite{Lu:2020bmd} and lensing searches of massive compact halo objects (MACHOs) towards the Large Magellanic Clouds (EROS)~\cite{Allsman:2000kg}, fast transient events near critical curves of massive galaxy clusters (Icarus)~\cite{Oguri:2017ock}, and observations of stars in the Galactic bulge by the Optical Gravitational Lensing Experiment (Ogle)~\cite{Niikura:2019kqi}. 

Finally, the NANOGrav experiment searching for a stochastic GW background in the frequency range close to $f \simeq 1\,{\rm yr}^{-1}$ would be able to detect the GWs induced at second order by the curvature perturbations responsible for PBH formation.
We show the constraint obtained by the null observation in the $11$-yr dataset
~\cite{Chen:2019xse}. We stress that this is only applicable for PBHs formed from the collapse of density 
perturbations and in the absence of non-Gaussianities (see~\cite{Nakama:2016gzw,Garcia-Bellido:2017aan,Cai:2018dig,Unal:2018yaa,Cai:2019elf}). The NANOGrav collaboration has recently released a new dataset based on $12.5$ yrs of observations, claiming that the previous constraint should relax due to an improved treatment of the intrinsic pulsar red noise~\cite{Arzoumanian:2020vkk}.
The collaboration also claims strong evidence for a stochastic common-spectrum process~\cite{Arzoumanian:2020vkk} in the new dataset, which could be explained by PBH formation \cite{DeLuca:2020agl,Vaskonen:2020lbd,Kohri:2020qqd,Domenech:2020ers,Sugiyama:2020roc} (although the signal could also be ascribed to supermassive BH binaries~\cite{Sesana:2004sp}, cosmic strings~\cite{Blasi:2020mfx, Ellis:2020ena, Buchmuller:2020lbh,Samanta:2020cdk}, phase transitions in a dark sector~\cite{Nakai:2020oit, Addazi:2020zcj,Ratzinger:2020koh}, or other scenarios~\cite{Namba:2020kij,Neronov:2020qrl,Li:2020cjj,Paul:2020wbz,Bhattacharya:2020lhc}).
As the new constraint would have a similar impact in the mass range of interest, here we choose to show the $11$-yr constraint as a reference~\cite{Chen:2019xse}. 

Notice that the bounds, typically derived for a monochromatic PBH population, can be adapted to extended mass functions using the techniques described in~\cite{Carr:2017jsz,Bellomo:2017zsr}. One should bear in mind the difference between constraints applying to high-redshift abundances and masses (such as Planck D/S and NANOGrav, which probe the early Universe physics) and the ones constraining late-time Universe quantities after the onset of  structure formation. The evolution of masses and $f_\PBH$ with accretion requires constraints to be treated as described in detail in Ref.~\cite{DeLuca:2020fpg}.  The main effect of accretion is to alleviate early Universe CMB constraints by shifting them to higher {\it late-time} mass ranges and making them weaker due to the growth of $f_\PBH$ [cf. Eq.~\eqref{fPBHevo}].

\subsection{The  GWTC-2 dataset confronts the PBH constraints}

In Fig.~\ref{fig:constraintsO1O2O3a} we collect all constraints on the PBH abundance and compare them to the population inferred from the GWTC-2 dataset.

At face value, if interpreted as coming from the PBH scenario, the GWTC-2 events seem to be
in tension with Planck~D.
However, one should consider this conclusion with a grain of salt. As already mentioned, the assumption of a thin disk in the Planck~D constraint is less reliable at high redshift~\cite{Ricotti:2007au} with respect to spherical accretion (Planck~S, which is compatible with GWTC-2).

We also take the opportunity to notice that the constraint from the NANOGrav $11$-yr data from Ref.~\cite{Chen:2019xse} has large systematic uncertainties, above all in their choice of the threshold $\zeta_c = 1$ for PBH formation (where $\zeta$ is the curvature perturbation responsible for the creation of PBHs upon collapse).
In order to account for these uncertainties, we have shown how the constraint is relaxed by choosing a  threshold $\zeta_c = 0.6$ motivated by state-of-art numerical simulations~\cite{Atal:2019erb} (see also the discussion in Ref.~\cite{Green:2020jor}).
Given the uncertainties discussed above, in Fig.~\ref{fig:constraintsO1O2O3a} we have decided to show the Planck~D and the more stringent NANOGrav constraints ($\zeta_c = 1$) by dashed lines without filling the corresponding excluded region.

In conclusion, assuming that all of the events in the GWTC-2 catalog are originated from PBHs is not in contrast with current observational constraints.

\section{Discussion}
\label{sec:Discussion}

This paper is a first step toward systematically testing various models for the formation of BH binaries.
We use a machine learning enhanced population analysis pipeline to constrain the PBH scenario with the latest GWTC-2 data. We find a preference for a scenario in which PBHs experience a phase of accretion before the reionization epoch and spin up. We also find that PBHs may form about $0.3 \%$ of the DM in the Universe. This abundance is still compatible with other constraints.

This work can also be considered as a proof of principle, which can be extended in various directions by relaxing some of the assumptions of our analysis.
We have assumed that every binary BH detection has a primordial origin, neglecting other formation channels.
This is obviously a very strong assumption.
In the future, we will mix the PBH simulation with different astrophysical populations, such as isolated and dynamically-formed binaries, to produce a more comprehensive inference model.
Furthermore, it would be interesting to extend the analysis to different PBH mass functions and accretion models.

For simplicity, we account for selection bias using a semianalytical noise model, which was checked against previous detection rate estimates.
However our single-detector approximation is expected to fail as the sensitivity improves and more detectors join the network.
Furthermore, we are using the signal-to-noise ratio instead of the false-alarm rate as our detection statistics, at variance with the LVC search pipelines.
Recent work uses machine learning techniques to better capture the detector network response~\cite{Wong:2020wvd,Gerosa:2020pgy}.
Future work should incorporate these techniques once they are validated against the results produced by a search pipeline, to better account for the selection bias.

\vspace{0.4cm}

\acknowledgements

\vspace{-0.2cm}

We thank T.~Helfer and D.~Gerosa for useful discussions.
Some computations were performed at the University of Geneva on the Baobab cluster.
E.~Berti, V.~Baibhav and K. W. K. Wong are supported by NSF Grants No. PHY-1912550 and AST-1841358, NASA ATP Grants No. 17-ATP17-0225 and 19-ATP19-0051, and NSF-XSEDE Grant No. PHY-090003. This work has received funding from the European Union’s Horizon 2020 research and innovation programme under the Marie Skłodowska-Curie grant agreement No. 690904. This research project was conducted using computational resources at the Maryland Advanced Research Computing Center (MARCC). 
V.DL., G.F. and A.R. are supported by the Swiss National Science Foundation 
(SNSF), project {\sl The Non-Gaussian Universe and Cosmological Symmetries}, project number: 200020-178787.
P.P. acknowledges financial support provided under the European Union's H2020 ERC, Starting Grant agreement no.~DarkGRA--757480, under the MIUR PRIN and FARE programmes (GW-NEXT, CUP:~B84I20000100001).
The authors would like to acknowledge networking support by the GWverse COST Action CA16104, ``Black holes, gravitational waves and fundamental physics.'' We acknowledge support from the Amaldi Research Center funded by the MIUR program ``Dipartimento di Eccellenza''~(CUP: B81I18001170001).

\bibliography{PBHO3}

\end{document}